\documentclass[amsmath,amssymb,showpacs,twocolumn,10pt]{revtex4}
\usepackage{graphicx}
\usepackage{psfig}

\newcommand{\s}{\ensuremath{\psi(t,r)}}
\newcommand{\n}{\ensuremath{\nu(t,r)}}
\newcommand{\T}{\ensuremath{\theta}}
\newcommand{\pt}{\ensuremath{p_\theta}}

\newcommand{\e}{Eq.$\;$} 
\newcommand{\M}{\ensuremath{{\cal M}}}
\newcommand{\prz}{\ensuremath{p_{r_0}}}

\newcommand{\ptz}{\ensuremath{p_{\theta_0}}}

\newcommand{\X}{\ensuremath{{\cal X}}}

\newcommand{\ti}{\mbox{$t_{i}$}}

\def\be{\begin{equation}}
\def\eq{\end{equation}}

\begin{document}
\preprint{}
\title{Gravitational collapse in asymptotically Anti-de Sitter/de Sitter 
backgrounds}
\author{T. Arun Madhav}
\email{arunmadhav@gmail.com}
\author{Rituparno Goswami}
\email{goswami@tifr.res.in}
\author{Pankaj S. Joshi}
\email{psj@tifr.res.in}
\affiliation{*Birla Institute of Technology and Science, Pilani \\\dag \ddag  
Tata Institute of Fundamental Research, Mumbai, India}

\begin{abstract}
{We study here the gravitational collapse of a matter cloud with a 
non-vanishing tangential pressure in the presence of a non-zero cosmological 
term. Conditions for bounce and singularity formation are 
derived for the model. It is also shown that when the tangential pressures 
vanish, the bounce and singularity conditions reduce to that of the dust 
case studied earlier. The collapsing interior is matched with an exterior which is asymptotically de Sitter or anti de Sitter, depending on the sign of cosmological constant. The junction conditions for 
matching the cloud to exterior are specified. The effect of the cosmological term on apparent horizons is studied in some detail, and the nature of central 
singularity is analyzed. We also discuss here the visibility of the singularity and implications for the cosmic 
censorship conjecture.}
\end{abstract}


\maketitle

\begin{section}{Introduction}
\par
The gravitational collapse of a matter cloud which is 
pressureless dust, and its dynamical evolution as governed by the 
Einstein's equations, were first studied 
in detail by Oppenheimer and Snyder
~\cite{kn:os}.
In recent years there have been extensive studies on gravitational 
collapse to examine the final fate of a collapsing cloud in
order to investigate the final state of such a collapse in terms
of formation of a black hole or naked singularity. These
studies would throw important light on the nature of cosmic censorship
and possible mathematical formulations for the same (for some
recent reviews, see e.g.
\cite{kn:penrose}).
Such studies have already helped to rule out several possible
versions of cosmic censorship, where a precise and
well-defined formulation itself has been a major unresolved problem so 
far. While understanding the nature of dynamical gravitational collapse 
within the framework of Einstein's gravity itself is a problem with
considerable astrophysical significance, the understanding
of cosmic censorship, if it is valid in some form, is another
major motivation for such collapse studies. 
Most of these investigations so far, however, have assumed a 
vanishing cosmological term ($\Lambda$).

The cosmological constant is sometimes thought of as a constant 
term in the Lagrangian density of general relativity, and it is also 
theorized that $\Lambda$ may be related to the energy density of vacuum (See
\cite{kn:smc} 
and references therein). Recent astronomical observations of high 
red-shift type Ia Supernovae
\cite{kn:pm} 
strongly indicate that the universe may be undergoing an accelerated 
expansion and it is believed that this may be due to a non-vanishing
\textit{ positive} cosmological constant. On the other hand, the proposed 
AdS/CFT conjecture
\cite{kn:mald} 
in string theory has generated interest in the possibility of 
space-times with a {\it negative} cosmological constant. The conjecture 
relates string theory in a spacetime where the non-compact part is 
asymptotically anti-de Sitter (AdS), to a conformal field theory in a 
space isomorphic to the boundary of AdS.  Whereas the obvious effect of 
a positive cosmological constant is to slow down gravitational collapse, 
that of the negative cosmological constant is to supplement the 
gravitational forces.

In this context, it seems pertinent to study the dynamical 
collapse of matter clouds when $\Lambda \neq 0$. Our purpose
here is to study a class of collapsing models which incorporate
pressure, and which are asymptotically either de Sitter
or anti-de Sitter geometry, depending on the sign of the cosmological
term. Dust models with 
$\Lambda$ are known in literature
\cite{kn:kra} 
and there have been some studies on dust collapse with $\Lambda$ in 
recent  years (see e.g.
\cite{kn:psj1}). 
We intend to discuss here a sufficiently general 
fluid model which allows pressure to be non-zero, and which also
allows the cosmological term to be non-vanishing. Specifically,
we study models with a non-vanishing {\it tangential pressure} $p_{\theta}$ 
\cite{kn:arp},
together with $\Lambda\neq0$. 
Collapse models with a tangential pressure have been studied
extensively, but not with a non-zero $\Lambda$
\cite{kn:tan}. 
Allowing the collapse to develop from regular initial
conditions, we study the bounce and singularity formation conditions, 
junction conditions at the boundary of the cloud so as to match it
to a suitable exterior, and we also consider how the 
apparent horizons are affected by the presence of $\Lambda$.  
One is also interested in considering the possibility 
whether the presence of $\Lambda$ could restore the {\it cosmic 
censorship conjecture} (see e.g. 
\cite{kn:gasp} 
and references therein), and to investigate how the final end 
state of gravitational collapse is affected by $\Lambda$. 
The present model generalizes the Lemaitre-Tolman-Bondi 
(LTB) dust 
\cite{kn:ltb}
model studies
with $\Lambda$, by introducing 
non-zero pressures in the collapsing cloud.

The relevant form of Einstein equations, conditions
to ensure that the collapse develops from a regular initial data, 
and the necessary energy conditions are introduced in Section II,
together with the details of the tangential pressure model.
In Section III, we derive the 
evolution of the collapsing matter shells, and explicitly give the 
conditions when a singularity is formed and 
when the bounce of a particular shell occurs during the collapse
evolution. The reduction to $\Lambda\neq0$ dust collapse case, 
when the tangential pressures are put to zero, is also demonstrated. 
For the collapsing solution to be physically plausible, 
it must satisfy certain junction conditions  at the 
boundary hypersurface where the interior collapsing cloud joins
with a suitable exterior spacetime. In 
Section IV we study the matching of the collapsing interior 
to the exterior Schwarzschild de Sitter or anti-de Sitter spacetime, 
and in Section V we discuss in 
brief the effect of $\Lambda$  on the  apparent horizons of 
the fluid model. In section VI the nature of the singularity in the tangential pressure fluid and dust models is considered when $\Lambda\neq0$, in terms of its being hidden within a black hole, or whether it would be 
visible to outside observers. The final Section VII summarizes 
some conclusions.

\section{Einstein Equations, Regularity and Energy conditions}

The general spherically symmetric metric in  
the comoving coordinates $(t,r,\theta,\phi)$ is given as,
\begin{equation}
ds^2= -e^{2\n}dt^2 + e^{2\s}dr^2 + R^2(t,r)d\Omega^2
\label{eq:metric}
\end{equation}
where $d\Omega^2=d\theta^{2}+sin^{2}\theta\,d\phi^{2}$ is the 
line element on a two-sphere. 
The metric variables $e^{2\n}$ and $e^{2\s}$ are functions
of $t$ and $r$. The energy-momentum tensor for a 
general {\it Type I} matter field
\cite{kn:hwe} 
can be written as,
\begin{equation}
T^t_t=-\rho;\; T^r_r=p_r;\; T^\T_\T=T^\phi_\phi=p_\T
\label{eq:setensor}
\end{equation}
The quantities $\rho$, $p_r$ and $p_\T$ are the matter density, 
radial pressure and the tangential pressure respectively. 
It is assumed that the matter field  satisfies the {\it weak energy 
condition}
\cite{kn:hwe}. 
It means that the energy density as 
measured by any local observer is positive. Then, for any timelike 
vector $V^\mu$ we have,
\begin{equation}
T_{\mu\gamma}V^\mu V^\gamma\ge0
\end{equation}
This amounts to,
\begin{equation}
\rho+\Lambda\ge0;\; \rho+p_r\ge0;\; \rho+p_\T\ge0
\end{equation}
The evolution of the matter cloud is determined by the 
Einstein equations and for the metric (\ref{eq:metric}) these are 
given by (in the units of $8\pi G=c=1$),
\be
\rho+\Lambda = \frac{F'}{R^2R'} , \; \; \; \;
  p_{r}-\Lambda =-\frac{\dot{F}}{R^2 \dot{R}}
\label{eq:t6}
\eq
\be
\nu'(\rho+ p_{r})=2(p_{\theta}-p_{r})\frac{R'}{R}-p_{r}'
\label{eq:t7}
\eq
\be
-2 \dot{R'}+R'\frac{\dot{G}}{G}+\dot{R}\frac{H'}{H}=0
\label{eq:t8}
\eq
\be
G-H=1 - \frac{F}{R}
\label{eq:t9}
\eq
where $(\,\dot{}\,)$ and $(')$ represent partial derivatives 
with respect to $t$ and $r$ respectively and,
\be
G(r,t)=e^{-2\psi}(R')^2, \; \; H(r,t)=e^{-2\nu}\dot{R}^2
\eq  
Here $F(r,t)$ is an arbitrary function, and in spherically symmetric 
spacetimes, it has the interpretation of the {\it mass function} of 
the collapsing cloud, in the sense that this represents the 
total gravitational mass within a shell of comoving radius $r$
\cite{kn:sharp}. 
The boundary of the collapsing cloud is labelled by the 
comoving coordinate $r_{\pi}$. In order to preserve  regularity at 
initial epoch we require $F(t_i,0)=0$.

 It can be seen from equation ($\ref{eq:t6}$) that the density of 
the matter blows up when $R=0$ or $R'=0$. The case $R'=0$ corresponds to  
{\it shell-crossing} singularities. The shell-cross singularities are
generally considered to be weak and possibly removable singularities.
Hence we shall consider here only the 
{\it shell-focusing} singularities (taking $R'>0$), where the 
physical radius of all the matter shells go to a zero value ($R=0$). Let 
us use the scaling independence of the coordinate $r$ to write,

\begin{equation}
R(t,r)=rv(t,r)
\label{eq:R}
\end{equation}
where `$v$' is the {\it Scale factor}. We have,
\begin{eqnarray}
v(t_i,r)=1; & v(t_s(r),r)=0; & \dot{v}<0
\label{eq:v}
\end{eqnarray}

where $t_i$ and $t_s$ stand for the initial and the singular 
epochs respectively. This means we scale the radial coordinate $r$ in 
such a way that at the initial epoch $R=r$, and at the singularity, $R=0$. 
The condition $\dot{v}<0$ signifies that we are 
dealing with gravitational collapse.  From the point of view of initial 
data, at the initial epoch $t=t_i$, we  have five functions of 
coordinate $r$  given by $\nu_0(r),\psi_0(r), \rho_0(r),p_{r_0}(r)$ \& 
$p_{\T_0}(r)$. Note that initial data are not all mutually 
independent. To preserve regularity and smoothness of initial data 
we must make some assumptions about the initial pressures at the regular 
center $r=0$. Let the gradients of pressures vanish at the center, 
that is, $\prz'(0)=\ptz'(0)=0$. The difference between radial and 
tangential pressures at the center should also vanish, 
i.e. $\prz(0)-\ptz(0)=0$. It is seen that we have a total of five 
field equations with seven unknowns, $\rho$, $p_r$, $\pt$, $\psi$, 
$\nu$, $R$, and $F$, giving us the freedom to choose two free functions. 
Selection of these functions, subject to the given initial data and weak 
energy condition, determines the matter distribution and metric 
of the space-time and thus leads to a particular dynamical evolution 
of the initial data.

Spherically symmetric collapse models, where the radial pressure 
is taken to be vanishing, but the tangential pressure could be non-zero 
have been studied in some detail over the past few years 
\cite{kn:tan}. 
The main motivation in the present consideration is to study bounce and 
singularity 
formation conditions for the case when we have a non-vanishing $\Lambda$
term present, when pressures are also introduced and allowed to be 
non-zero 
within a collapsing cloud. One would also like to understand how 
$\Lambda$ affects the junction conditions at the boundary where the cloud 
is matched to an exterior spacetime, the horizon 
formation, and the nature of central singularity, when non- 
zero pressures are present within the cloud.

Taking $p_{r}=0$ in Eq. (\ref{eq:t6}) gives $F(t,r)$ and $\rho$ 
in the following forms,
\begin{equation}
F(t,r)=r^3\M (r) + \frac{\Lambda}{3} R^{3}
\label{eq:mass}
\end{equation}
\begin{eqnarray}
\rho=\frac{3\M+r\left[\M_{,r}\right]}{v^2(v+rv')}
\label{eq:rho} 
\end{eqnarray}
Here $\M (r)$ is an arbitrary function of $r$ subject to the energy 
conditions. 
There remains the freedom to choose one function, since there are 
six equations with seven unknowns. In order to work within the 
framework of a specific class of models, we 
take $\n$ in the specific form,
\begin{equation}
\n=c(t)+\nu_0(R)
\label{eq:nu}
\end{equation}

The conditions imposed here, namely that of vanishing radial pressures, 
and Eq. (\ref{eq:nu}) specifying a form of $\nu$ may be considered 
to be strong assumptions. However, this enables us to make our study in 
sufficient generality with sufficiently rich structure as we shall
see below,
with non-zero pressures introduced into the collapse model. A mechanism 
by which we can have non-vanishing tangential pressures is illustrated by 
the Einstein cluster
\cite{kn:ein}. 
This is a spherically symmetric cluster of 
rotating particles where the motion of the particles is sustained 
by an angular 
momentum. This has an average effect and creates non-zero 
tangential stresses within the collapsing cloud.

Also, one can rederive the dust collapse models with a non-vanishing
cosmological term,  
by putting $\nu_0(R)=0$ in \e (\ref{eq:nu}) and 
redefining the comoving time coordinate. 
It is thus clear that the class of models considered generalize the 
dust collapse models with a non-vanishing cosmological constant.
In general, 
as $v\rightarrow 0$, 
$\rho\rightarrow\infty$. Thus the density blows up at the singularity 
$R=0$ which will be a curvature singularity as expected. 
Using \e (\ref{eq:nu}) in \e (\ref{eq:t8}), we have,
\begin{equation}
G(t,r)=b(r)e^{2\nu_0(R)}
\label{eq:G}
\end{equation}
Here $b(r)$ is an arbitrary, $C^{\,2}$ function of $r$. 
In correspondence with the dust models, we can write,
\begin{equation}
b(r)=1+r^2b_0(r)
\label{eq:veldist}
\end{equation}
where $r^{2}b_0(r)$ is the {\it energy distribution function} 
for the collapsing shells. Using the given initial data and \e(\ref{eq:t7}) 
one can obtain the function $\nu_{0}(R)$. Substituting \e(\ref{eq:nu}) 
in \e(\ref{eq:t7}) we get the intrinsic equation of state,
\begin{equation}
\pt=\frac{R}{2}\nu_{,R}\,\rho
\label{eq:ptheta}
\end{equation}

Finally, using equations 
(\ref{eq:mass}),(\ref{eq:nu}) and (\ref{eq:G}) in \e(\ref{eq:t9}), we have,
\begin{equation}
\sqrt{R}\dot{R}=-a(t)e^{\nu_0(R)}\sqrt{(1+r^2b_0)Re^{2\nu_0}-R+r^3\M
+\frac{\Lambda}{3}R^{3}}
\label{eq:collapse}
\end{equation}
Here $a(t)$ is a function of time and by a suitable scaling of the time 
coordinate, we can always make $a(t)=1$. We deal here with the 
collapse models, and so the negative sign is due to the 
fact that $\dot{R}<0$, which represents the collapse condition. 
\end{section}

\begin{section}{Singularity formation and Rebounce}

One of our main purposes here is to examine how the introduction
of a non-vanishing cosmological term modifies the collapse dynamics.
For example, in the case of dust collapse, once the collapse initiates
from an initial epoch, there cannot be any reversal or a bounce, and
the gravity forces the cloud to collapse necessarily to a singularity.
However, this need not be the case when the cosmological term is
non-vanishing, and we have to re-examine the collapse dynamics in order
to find how the collapse evolves. This we do here for the particular
class of tangential collapse models as specified above, which also 
generalizes the dust collapse case.

The evolution of a particular shell may be deduced from 
\e (\ref{eq:collapse}). Rewriting \e (\ref{eq:collapse}) in terms of 
the scale factor we have,
\be
\dot{v}^{2}= \frac{e^{2\nu_{0}}\,\,[v\,j(r,v)+\M+\frac{\Lambda}{3}v^{3}]} 
{v}={V}(r,v)
\label{eq:gendyn}
\eq
where
\be
j(r,v)=\frac{b(r)e^{2\nu_{0}(rv)}-1}{r^{2}}
\eq
The right hand side of the equation (\ref{eq:gendyn}) may be thought of as 
an 
{\it effective potential} (${V}(r,v)$) for a shell. The allowed regions of 
motion correspond to ${V}(r,v)\geq0$, as $\dot{v}^{2}$ is non-negative, 
and 
the dynamics of the shell may be studied by finding the turning points. 
If we start from an 
initially collapsing state ($\dot{v}<0$), we will have a rebounce if 
we get  $\dot{v}=0$, before the shell has become singular. This can 
happen when ${V}(r,v) = 0$. Hence, to study the various evolutions for a 
particular shell we must analyze the roots of the equation 
${V}(r,v) = 0$ keeping the value $r$ to be fixed. It will be seen that the 
cosmological constant appearing 
in the effective potential does play an important role in the evolution 
of a shell.

To clarify these ideas, let us consider a smooth initial data, 
where the initial density, pressure, 
and energy distributions are expressed as only even powers of $r$. 
Such a consideration, that the initial data be smooth, is often
justified on physical grounds. So we take,
\be
\rho(\ti,r)=\rho_{00}+\rho_{2}r^2 +\rho_{4}r^4+\cdots
\eq
\be
p_{\theta}(\ti,r)=p_{\theta_2}r^2 +p_{\theta_4}r^4+\cdots 
\eq
\be
b_0(r)=b_{00}+b_{02}r^2+\cdots
\eq 
With the above form of smooth initial data to evolve in time
using the Einstein equations, we 
can explicitly 
integrate  \e (\ref{eq:t7}) at the initial epoch to get,
\be
\nu_{0}(R)=p_{\theta_2} R^2 +\frac{(p_{\theta_4}-\rho_{2}p_{\theta_2})}
{2} R^4+\cdots
\label{eq:nu1}
\eq
We have neglected here higher order terms in the expansion, since at 
present we want to 
concentrate on the evolution of shells near $r=0$. The conditions when 
the treatment is applicable to the whole cloud will be discussed later. 
\e (\ref{eq:gendyn}) near the center of the cloud 
\hbox{($r<<r_{\pi}$)} may be written as,
\be
\dot{v}^{2}= \, \frac {(1+2p_{\theta_2}r^{2}v^{2})\,[\,(\,\Lambda+6p_
{\theta_2}b(r)\,)\,v^{3} + 3b_{0}(r)\,v + 3\M]}{3v}
\label{eq:dyn}
\eq
The first factor in $V(r,v)$ is initially positive, because it 
is the $|g_{00}|$ term. As the collapse evolves, the scale 
factor ($v$) reduces from $1$ at the initial epoch to $0$  at the time 
of singularity. Hence it clear that the first factor can never become zero, 
and hence does not contribute to a bounce of the shell. The main 
features of the evolution of the cloud basically derive 
from the second factor in $V(r,v)$.

The second factor in the effective potential expression is 
a cubic equation which in general has three roots. Only positive real 
roots correspond to physical cases. Since 
the coefficient in the second power is zero,we may conclude that if 
all three roots are real then at least one of them has to be positive 
and at least one negative. We observe that 
$V(r,0)=\mathcal{M}>0$. Hence, any region between $R=rv=0$  and the first 
positive zero 
of $V(r,v)$ always becomes singular during collapse. The region between 
the unique positive roots is forbidden since in those regions 
$\dot{v}^{2}<0$. For a particular shell to bounce 
it must therefore lie, during initial epoch ($v=1$), in a region to 
the right of the second positive root. We will now analyze the various 
cases for $\Lambda\neq0$ in detail and derive the necessary conditions.

1. If 
\be
b_{0}(r)\geq 0 \,\, ; \,\,
(\,{\Lambda}+6p_{\theta_2}b(r)\,)>0
\eq
then from the Descartes's rule of signs (See for example
\cite{kn:peq}) 
we see that there are no positive roots. 
Thus a {\it singularity always forms} from initial collapse.

2. If
\be
b_{0}(r)\geq 0 \,\, ; \,\,
(\,{\Lambda}+6p_{\theta_2}b(r)\,)<0
\eq
 we infer from the sign rule that there is exactly one 
positive root $\left(\alpha(r)\right)$. The other two roots are negative 
or complex conjugates. 
The allowed space of dynamics is $[0,\alpha]$. Thus $\alpha\ge1$ 
would always ensure a singularity. However $\alpha<1$ implies an 
unphysical situation 
initially, where $\dot v^{2}<0$.

3. If
\be
b_{0}(r)\leq 0 \,\, ; \,\,
(\,{\Lambda}+6p_{\theta_2}b(r)\,)<0
\eq
there is exactly one positive root $\left(\beta(r)\right)$. 
Again all shells in the allowed dynamical space $[0,\beta]$ become 
{\it singular} starting from initial collapse.

4. If
\be
b_{0}(r)< 0 \,\, ; \,\,
(\,{\Lambda}+6p_{\theta_2}b(r)\,)>0
\label{eq:mainbounce}
\eq
there are three possibilities,
\par
4.1 If
\be
\mathcal{M}^{2}\,>\,-4\,\frac{b^{3}_{0}(r)}{9({\Lambda}+6p_{\theta_{2}}b(r))}
\eq
then there are no positive roots and a {\it singularity} is 
always the final outcome of collapse for shells under consideration.
\par
4.2 If
\be
\mathcal{M}^{2}\,<\,-4\,\frac{b^{3}_{0}(r)}{9({\Lambda}+6p_{\theta_{2}}b(r))}
\label{eq:bounce}
\eq
then there are two positive roots ($\gamma_{1}(r)$ and $\gamma_{2}(r)$ 
respectively). 
The space of allowed dynamics is $[0,\gamma_{1}]$ and $[\gamma_{2},\infty)$. 
The region $(\gamma_{1},\gamma_{2})$ is forbidden. Shells in the 
$[0,\gamma_{1}]$  
region initially, {\it always become singular}. But shells initially 
belonging to the region 
$[\gamma_{2},\infty)$ will {\it undergo a bounce and subsequent expansion} 
starting from initial collapse. This bounce occurs  
when their geometric radius approaches 
$R_{bounce}=r\gamma_{2}$.
Using the definitions
\be
\varrho=\sqrt{\frac{-4b_{0}(r)}{\Lambda+6p_{\theta_{2}}b(r)}}
\eq
\be
\vartheta=\,\frac{1}{3}\, Cos^{-1}\left[-\sqrt{\frac{-9\mathcal{M}(r)^{2}
(\Lambda+6p_{\theta_{2}}b(r))}{4b_{0}^{3}(r)}}\right]
\eq
the condition for a particular shell to become singular or undergo 
a bounce may be 
explicitly written in terms of the initial data and $\Lambda$ as,
\be
1<\varrho\,\,Cos\vartheta \,\,;\,\,\,\, Singularity
\eq
\be
1>\varrho\,\,Cos(\vartheta+\frac{4\pi}{3}) \,\,;\,\,\,\, Bounce
\label{eq:bounceshell}
\eq
Here $\varrho\,\,Cos\vartheta$ is $\gamma_{1}(r)$ and 
$\varrho\,\,Cos(\vartheta+\frac{4\pi}{3})$ is $\gamma_{2}(r)$, 
which are the 
two roots of the potential function. Note that contrary to the dust models 
with $\Lambda$, there may be a bounce for both positive and negative 
values of the cosmological constant.
\par
4.3 If 
\be
\mathcal{M}^{2}\,=\,-4\,\frac{b^{3}_{0}(r)}{9({\Lambda}+6p_{\theta_{2}}b(r))}
\eq
the positive roots are equal. There is no forbidden region, and there will 
be a bounce
if $1>\frac{\varrho}{2}$.
Fig. 1. illustrates a particular choice of initial data and $\Lambda$ which 
causes a bounce in the central ($r=0$) shell. 
\begin{figure}[t!!]
\centering
\includegraphics[width=5.75cm,angle=-90]{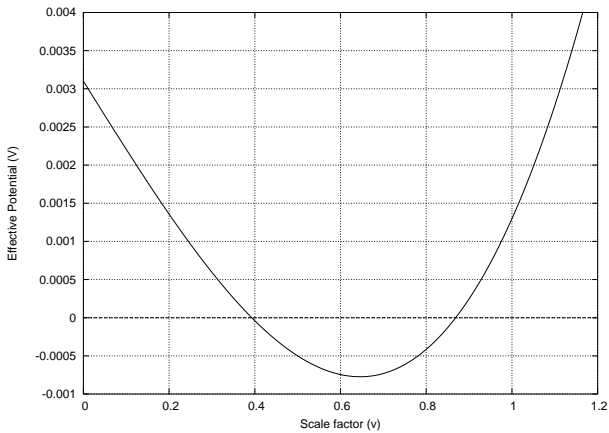}
\caption{The effective potential profile (V(r,v)) for the $r=0$ shell with 
$(\Lambda\,+\,6\,p_{\theta_{2}})=7.1978\times10^{-3}\,\,, b_{00}=-3\times10^
{-3}\,$, and $\,\,\mathcal{M}=1.033\times10^{-3}$ in appropriate units.}
\end{figure}

To analyze the evolution of shells far from the center, 
in general one has to resort to numerical methods, and it is  
difficult to analytically give a simple expression for 
the singularity or bounce conditions. Nevertheless, the analysis as 
given here becomes 
valid for the entire cloud all the way till boundary when the geometrical 
radius of the cloud boundary at the 
initial epoch ($R\left(t_{i},r_{\pi}\right)=r_{\pi}$) is itself 
small relative to the initial data coefficients (ie. $\rho_{n}r_{\pi}^{n},
p_{\theta_{n}}r_{\pi}^{n}<<1$). Also, if we choose the initial data 
such that the higher coefficients in the power series expansion are zero 
(i.e. $p_{\theta_{n\geq4}},\rho_{k\geq4}=0$) and 
$\rho_{2}r_{\pi}^{2},p_{\theta_{2}}r_{\pi}^{2}<<1$, then the analysis is 
again 
applicable to the whole cloud. Thus the results derived can be 
considered quite general in these circumstances and applicable to the 
cloud as a whole.

In this context, if all shells in the collapsing cloud satisfy 
\e(\ref{eq:mainbounce}), 
(\ref{eq:bounce}) and (\ref{eq:bounceshell}), the complete cloud undergoes 
bounce starting from initial collapse. To avoid shell crossings 
the sufficient condition would be,
\be
\forall\,\,\, r\,\,\, \epsilon\,\,\, [0,r_{\pi})\,\,\,\,\, ,\,\,\,\,\, 
\gamma_{2}(r+\delta)\geq\gamma_{2}(r)
\eq
where $\delta$ is an infinitesimal increment in the comoving radius. 
It is seen that in all the cases discussed, it is not $\Lambda$ alone, but 
${\Lambda+6p_{\theta_{2}}b(r)}$ along with $b_{0}(r)$ that determines the 
evolution of the shell. This is in contrast to the dust models with 
a non-zero $\Lambda$, 
where solely the cosmological constant decided the evolution of a shell 
for a given 
energy function. It is also interesting to note that unlike the dust 
collapse models with $\Lambda$, there could be a bounce in the 
fluid model with vanishing radial pressures for both positive and 
negative values of the cosmological constant. This is due to the 
contribution from the tangential pressure.

It can be seen now that one can rederive the known bounce conditions 
in $\Lambda\neq0$ dust collapse case (Deshingkar {\it et al}
\cite{kn:psj1}) 
as a special case of the consideration here, when $p_{\theta}=0$. 
In that case,
\be
\dot{v}^{2}=\,\frac{\Lambda v^{3}+3b_{0}(r)v+3\mathcal{M}}{3v}={V(r,v)}
\eq
The study now becomes valid for all shells, without any approximation. 
For example, 
following similar steps as for the fluid model, one obtains 
for $\Lambda>0$,\\

1. If
 \be
  b_{0}(r)>0 
  \eq
the singularity always forms from initial collapse.\\

2. If
\be
 b_{0}(r)=0 
\eq
then again it is found that all shells become singular from collapse.\\

3. If
\be
 b_{0}(r)<0 
\eq
there are two scenarios possible. For 
\be
\mathcal{M}^{2}\,>\, -4\,\frac{b^{3}_{0}}{9\Lambda}
\eq
all shells become singular.
For
\be
\mathcal{M}^{2}\,<\, -4\,\frac{b^{3}_{0}}{9\Lambda}
\eq 
there are two positive roots ($\psi_{1}$ \& $\psi_{2}$ respectively) 
for $V(r,v)$. 
Shells belonging to $[0,\psi_{1}]$ always become singular while 
those belonging to 
$[\psi_{2},\infty)$ undergo bounce starting from initial collapse.
\end{section}

\begin{section}{Spacetime matching}

As we pointed out above, there is a strong physical motivation 
to study and investigate the gravitational collapse phenomena in a 
background which is asymptotically either a de Sitter or anti-de Sitter 
metric. For this purpose, the collapsing cloud has to be matched at the 
boundary to a suitable exterior spacetime which has the desired 
properties.

In the present case, we shall show below that the exterior vacuum 
spacetime of the collapsing region may be described by the Schwarzschild- 
de Sitter (SdS) or Schwarzschild- Anti de Sitter (SAdS) metric, 
depending on whether the cosmological constant is taken to be positive or 
negative. The collapsing interior cloud which has a non-zero tangential
pressure is then to be smoothly matched to an exterior spacetime in
order to 
generate the full spacetime. The necessary and sufficient conditions 
to achieve a smooth matching are given by the Israel-Darmois junction 
conditions 
(\cite{kn:dar},\cite{kn:bonnor}), 
which we shall use below.

Let the interior of the collapsing cloud be described by the metric,
\begin{equation}
S^{-}: ds^2_{-}= -e^{2\n}dt^2 + e^{2\s}dr^2 + R^2(t,r)d\Omega^2
\label{eq:m1}
\end{equation}
The exterior vacuum solution can be given as 
the Schwarzschild-de Sitter/Anti de Sitter spacetime (as decided
by the sign chosen for the cosmological term) as given by,
\be
S^{+}: ds^2_{+}= -\mathcal{D}(\mathcal{R})dT^2 + \mathcal{D}^{-1}
(\mathcal{R})\,d\mathcal{R}^2 + \mathcal{R}^2d\Omega^2
\label{eq:m2}
\eq
where $\mathcal{D}(\mathcal{R})$ is given by,
\be 
\mathcal{D}(\mathcal{R})= 1-\frac{2M}{\mathcal{R}}\mp\frac
{|\Lambda| \mathcal{R}^{2}}{3}
\eq 
The negative sign precedes $|\Lambda|$ for a SdS exterior and the 
positive for SAdS exterior. 
Note that as $\mathcal{R}\rightarrow\infty$, the $\mp\frac{|\Lambda| \mathcal
{R}^{2}}{3}$ 
term dominates over the $\frac{2M}{\mathcal{R}}$ term, and the spacetime 
approaches asymptotically 
de Sitter metric ($-\frac{|\Lambda| \mathcal{R}^{2}}{3}$), or the Anti-de 
Sitter metric
($+\frac{|\Lambda| \mathcal{R}^{2}}{3}$), as the case may be.

Let $\Pi$ denote the boundary hypersurface. The equations of 
the boundary hypersurface considered as an embedding in the interior 
or exterior spacetimes are,
\be
\Pi^{-} : r-r_{\pi}=0\,\,\,\,\,\,\,\,
\Pi^{+} : \mathcal{R}-\mathcal{R}_{\pi}(T)=0
\label{eq:s1}
\eq
Substituting (\ref{eq:s1}) in (\ref{eq:m1}) and (\ref{eq:m2}) 
we get the metric on the hypersurface as,
\be
S^{-}_{\Pi}: ds^2_{-}= -e^{2\nu(t,r_{\pi})}dt^2 + R^2(t,r_{\pi})d\Omega^2 
\label{eq:h1}
\eq

\be
S^{+}_{\Pi}: ds^2_{+}= -\mathcal{D}(\mathcal{R}_{\pi})dT^2 
+ \mathcal{D}^{-1}(\mathcal{R}_{\pi})\,d\mathcal{R}^2_{\pi} 
+ \mathcal{R}_{\pi}^2 d\Omega^2
\label{eq:h2}
\eq

The Israel-Darmois conditions to match the interior spacetime
with exterior require that 
the first and second fundamental 
forms of the boundary hypersurface match. The first fundamental form 
is given by,
\be 
g_{\mu \nu}\,d\zeta^{\mu}d\zeta^{\nu}
\eq
where $\zeta$ parametrizes the hypersurface 
($\zeta^{\iota}:\tau,\theta,\phi$). The matching of the first fundamental 
form gives, from \e (\ref{eq:h1}) and \e (\ref{eq:h2}),

\be
R(t,r_{\pi})=\mathcal{R}_{\pi}
\label{eq:f1}
\eq

\be
d\tau=e^{\nu(t,r_{\pi})} dt
\label{eq:f2}
\eq

\be
d\tau=\left[\mathcal{D}_{\pi} - \frac{{\mathcal{R}}_{\pi\, ,T}^{2}}
{\mathcal{D}_{\pi}}\right]^{1/2} dT
\label{eq:f3}
\eq
The above three conditions must be satisfied for a smooth matching of the 
collapsing interior to the exterior spacetime. The next set of conditions 
will be given by the matching of the second fundamental forms.

The second fundamental form is given by 
\be
K_{\mu\nu}\,d\zeta^{\mu}d\zeta^{\nu}
\eq
where, 
\be
K_{\mu\nu}^{\pm}=-n_{\sigma}^{\pm} x^{\sigma}_{,\zeta^{\mu},
\zeta^{\nu}}-n_{\sigma}^{\pm}\Gamma^{\sigma}_{\beta\gamma}x^{\beta}_
{\, ,\zeta^{\mu}}x^{\gamma}_{\, ,\zeta^{\nu}} 
\eq
is the extrinsic curvature
\cite{kn:israel}. 
The (,) denotes partial differentiation. $n_{\sigma}$ is the normal 
to the hypersurface which is given by,
\be
n_{\sigma}=\frac{f_{,\sigma}}{[ 
g^{\mu\nu}f_{,\mu}f_{,\nu}]^{1/2}}
\eq
where $f=0$ is the equation of the boundary hypersurface. 
Direct calculation gives,
\be
n_{\sigma}^{-}=(0,e^{\psi(t,r_{\pi})},0,0)\,\,\,\,\,\,
\eq
\be
n_{\sigma}^{+}=(-\mathcal{R}^{\pi}_{,\tau}\, ,T_{,\tau}\, ,0\, ,0)
\eq
The $K_{\theta\theta}$ extrinsic curvatures are calculated as,
\be
K^{+}_{\theta\theta}=\,\,(\mathcal{D}\mathcal{R}T_{,\,\tau})_
{\pi}\,\,\,\,\,\,\,K^{-}_{\theta\theta}=\,\,(R R_{,r}e^{-\psi})_{\pi}
\label{eq:mktt}
\eq
Now using the fact that the second fundamental forms match 
(i.e. $[K_{\theta\theta}^{+}-K_{\theta\theta}^{-}]_{\pi}=0$), we get using 
equations (\ref{eq:f1}),(\ref{eq:f2}),(\ref{eq:f3}) and (\ref{eq:mktt}) 
after simplification,
\be
R_{\pi\, , r}^{2}\,e^{-2\psi} - R_{\pi\, ,\tau}^{2}\,e^{-2\nu}
=\,\,\,1-\frac{2M}{\mathcal{R}_{\pi}}\mp\frac{|\Lambda|}{3}\mathcal
{R}_{\pi}^{2}
\label{eq:mc}
\eq
This is identical to the Cahill and McVittie definitions for the mass
function 
(\cite{kn:sharp},\cite{kn:kmc}), 
and hence from \e (\ref{eq:t9}) we may by comparison take
\be
F_{\pi} = 2M\pm\frac{|\Lambda|}{3}\mathcal{R}_{\pi}^{3}
\label{eq:mas} 
\eq
This expression suggests that for a smooth matching of the interior and 
exterior spacetimes the interior mass function at the surface must equal 
the generalized Schwarzschild mass.

As we can see there is a contribution to the mass function from the 
cosmological 
constant. A positive cosmological constant has an additive contribution 
and a negative cosmological constant has a deductive contribution 
to the mass function $F_{\pi}$. The matching of a collapsing dust interior 
with exterior SdS/SAdS would give similar results 
\cite{kn:psj1}). 
The $K_{\tau\tau}$ and $K_{\phi\phi}$ components of the extrinsic 
curvature may similarly be calculated. The condition 
$[K_{\tau\tau}^{+}-K_{\tau\tau}^{-}]_{\pi}=0$ gives no new information 
since the radial pressure is zero. Due to spherical symmetry, 
$[K_{\phi\phi}^{+}-K_{\phi\phi}^{-}]_{\pi}=0$ gives the same result 
as \e (\ref{eq:mas}). All the above conditions {\it  (\ref{eq:f1}), 
(\ref{eq:f2}), (\ref{eq:f3}) and (\ref{eq:mas}) must be satisfied for 
smooth matching of spacetimes} across the boundary.
\end{section}

\begin{section}{Horizons}
{\it Apparent horizons} ($\mathcal{H}$) are the boundaries of 
trapped regions \cite{kn:hwe} in the spacetime. We discuss below the 
effect of a non-zero $\Lambda$ term on the 
apparent horizons of the tangential pressure fluid collapse models considered 
here briefly. In general, the equation of $\mathcal{H}$ can be 
written as,
\be
\mathcal{H} \,:\,\,\,\,\,g^{\mu\nu}\,R_{\, ,\mu}\,R_{\, ,\nu}=0
\label{eq:hor}
\eq
Substituting (\ref{eq:m1}) in (\ref{eq:hor}) we get,
\be
R_{\, , r}^{2}\,e^{-2\psi} - R_{\, ,\tau}^{2}\,e^{-2\nu}\,=\,0
\label{eq:hor1}
\eq
From the definition of the mass function  \e (\ref{eq:t9}), and 
(\ref{eq:hor1}) we therefore have,
\be
\,1\,\,-\,\,\frac{F}{R} \,=\,0
\label{eq:apph}
\eq
Finally, from (\ref{eq:mass}) and (\ref{eq:apph}),
\be
\mathcal{H}:\,\,\,(3-\Lambda\,R^{2})\,R=3\,r^{3}\mathcal{M}
\label{eq:aph}
\eq
\par
When $\Lambda=0$ there is necessarily only one apparent horizon 
given by 
\be
R=r^{3}\mathcal{M}
\eq
which is the Schwarzschild horizon in case we are considering
that geometry. The same equation also defines horizon
within the collapsing cloud, where $R(t,r)$ is one of the 
metric functions.
For the case when $\Lambda>0$, \e(\ref{eq:aph}) is a cubic 
equation with at least one 
positive and one negative root. For $\Lambda>0$ the various cases are
given as below.

1. For 
\be
3\,r^{3}\mathcal{M}\,<\,\frac{2}{\sqrt{\Lambda}}
\eq
there are two positive roots for (\ref{eq:aph}) and hence there 
are {\it two} apparent horizons. These horizons are given by
\begin{equation}
R_{c}(r)=\frac{2}{\sqrt{\Lambda}} Cos\left[\frac{1}{3} Cos^{-1}
(-\frac{3}{2}\,r^{3}\mathcal{M}\sqrt{\Lambda})\right]
\end{equation}
\begin{equation}
R_{b}(r)=\frac{2}{\sqrt{\Lambda}} Cos\left[\frac{4\pi}{3}
+\frac{1}{3} Cos^{-1}(-\frac{3}{2}\,r^{3}\mathcal{M}\sqrt{\Lambda})\right]
\end{equation}
These have been at times called the cosmological, and the black hole 
horizons 
\cite{kn:hsn}.

2. For
\be
3\,r^{3}\mathcal{M}\,=\,\frac{2}{\sqrt{\Lambda}}
\eq
there is only one positive root for (\ref{eq:aph}), given by
\begin{equation}
R_{bc}(r)=\frac{1}{\sqrt{\Lambda}}
\end{equation}
This corresponds to a {\it single apparent horizon}.

3. For 
\be
3\,r^{3}\mathcal{M}\,>\,\frac{2}{\sqrt{\Lambda}}
\label{eq:limit}
\eq
there are no positive roots and hence there are {\it no apparent horizons}.

The case (\ref{eq:limit}) also shows that the  mass of the 
black hole is bounded above by $F=\frac{1}{\sqrt{\Lambda}}$ and attains 
the largest proper area ${4\pi}/{\Lambda}$. A general result exists 
in literature
\cite{kn:hsn} 
showing that in spacetimes with $\Lambda>0$ and matter satisfying 
the strong energy condition, the area of a black hole cannot 
exceed ${4\pi}/{\Lambda}$. A detailed treatment of apparent horizons 
in the Lemaitre-Tolman-Bondi dust collapse models with a non-zero
$\Lambda$ term is given by Cissoko M. {\it et al}
\cite{kn:psj1}.

From \e(\ref{eq:collapse}), the time of apparent horizon 
formation in the fluid model is given by
\begin{equation}
t_{ah}(r)=t_{s}(r)-t_{1}(r)
\eq
where $t_{1}(r)$ is defined as,
\be
t_{1}(r)=\int_0^{v_{ah}(r)}\frac{\sqrt{v}dv}{\sqrt{e^{4\nu_0}vb_0+e^{2\nu_0}
\left(v^3h(rv)+\M + \frac{\Lambda}{3} v^{3} \right)}}
\label{eq:t1def}
\end{equation}
Also $rv_{ah}=R_{b}$ or $R_{c}$, $h(R)= (e^{2\nu_0(R)}-1)/R^2$ and 
$t_{s}(r)$ is the time of singularity formation. For shells close to the 
center of the cloud ($r<<r_{\pi}$) the expression becomes,
\be
t_{ah}(r)=t_{s}(r)-\int_0^{v_{ah}(r)}\frac{\sqrt{3v}dv}{\sqrt{\,(\,
\Lambda+6p_{\theta_2})\,v^{3} + 3b_{0}(r)\,v + 3\M}}
\label{eq:hcent}
\eq
 It is observed that the cosmological constant modifies the 
time of {\it formation of horizons} and also the  {\it time lag} between 
horizon formation and singularity formation in the fluid model.
\end{section}

\begin{section}{Nature of the Central Singularity}

The final end state of gravitational collapse and the nature of 
the resulting singularity continue to be among the most outstanding problems 
in gravitation theory and relativistic astrophysics today. 
As pointed out earlier, the hypothesis that such a collapse leading to a 
singularity, under 
physically realistic conditions must end in the formation of a black hole, 
and that the eventual singularity must be hidden below the event 
horizons of gravity 
is the {\it cosmic censorship conjecture}. 
Despite numerous attempts, this conjecture as such remains a major 
unsolved problem lying at the foundation of black hole physics today.

From such a perspective, we need to examine the nature of the
singularity, in terms of its visibility or otherwise for outside 
observers, when it develops within the context of the models considered
here. This should tell us how the presence of the $\Lambda$ term
modifies these considerations, because we already know that in the
case of a tangential pressure present, but a vanishing cosmological
constant, both black holes and naked singularities do develop as final
collapse end states depending on the nature of the initial data (see e.g.
\cite{kn:arp}).

We have already seen above that the cosmological constant modifies the 
time of formation of trapped surfaces. If the formation of the horizon 
precedes the formation of the central singularity then the singularity 
will be necessarily covered, i.e. it is a black hole. If on the other 
hand, the horizon formation occurs after the singularity formation, 
there may be future directed non-spacelike geodesics that end in the past 
at the singularity. Then the final end state would be a naked singularity. 
Thus we need to find whether there exist future directed null geodesics 
that end at the singularity in the past.

Towards analysing this issue,
let us define a function $h(R)$ as,
\begin{equation}
h(R)=\frac{e^{2\nu_0(R)}-1}{R^2}=2g(R)+{\cal O}(R^2)
\label{eq:h}
\end{equation}
Using \e(\ref{eq:h}) in \e(\ref{eq:collapse}), we get after 
simplification,
\begin{equation}
\sqrt{v}\dot{v}=-\sqrt{e^{4\nu_0}vb_0+e^{2\nu_0}\left((\frac{\Lambda}{3}
+h(rv))v^{3}+\M\right )}
\label{eq:collapse1}
\end{equation}
Integrating the above equation, we get,
\begin{equation}
t(v,r)=\int_v^1\frac{\sqrt{v}dv}{\sqrt{e^{4\nu_0}vb_0+e^{2\nu_0}
\left(v^3(h(rv)+\frac{\Lambda}{3})+\M\right)}}
\label{eq:scurve1}
\end{equation}
The time of formation of a shell focussing singularity, for a specific 
shell, is obtained by taking the limits of integration in above as (0,1). 
The shells collapse consecutively, one after the other to the center as 
there are no shell-crossings ($R'>0$). We are interested in the central 
shell (i.e. the singularity forming at $r=0$), since we will see 
that all $r>0$ shells are necessarily covered on becoming singular. 
Taylor expanding the above function around $r=0$, we get, 
\be 
t(v,r)=t(v,0)\;+\left.r\;\frac{d t(v,r)}{dr}\right|_{r=0}  
+\left.\frac{r^2}{2!}\;\frac{d^{2}t(v,r)}{d^2{r}^2}\right|_{r=0}+...
\label{eq:scurve2}
\eq
Let us denote, 
\be
\X_{n}(v)=\left.\frac{{d} ^{n} t(v,r)}{{d} r^{n}}\right|_{r=0}  
\eq
The initial data is taken to be smooth (ie. with only even powers 
of $r$ allowed). Due to this choice, the first derivative of the functions 
appearing in the above equation vanish at $r=0$. Hence we have,
\be
\X_{1}(v)=0
\eq
Defining $B_{f}(r,v)={e^{4\nu_0}vb_0+e^{2\nu_0}\left(\,\,
(\frac{\Lambda}{3}+h)v^{3}+\M\,\,\right )}$ we may write the Taylor 
expansion about $r=0$ as,
\be
t(v,r)=t(v,0)-\frac{r^{2}}{4}\int_{v}^{1}\frac{B^{''}_{f}(0,v)\sqrt{v}\,dv }
{B_{f}(0,v)^{3/2}}+...
\eq
and we have,
\be
\X_{2}(v)=\,\,-\int_{v}^{1}\frac{B^{''}_{f}(0,v)\sqrt{v}\,dv }
{B_{f}(0,v)^{3/2}}
\label{eq:ttan}
\eq

In  order to consider the possibility of existence of null geodesic 
families which end at the singularity in the past, and to examine 
the nature of the singularity occurring at $R=0, r=0$ in this model, let 
us consider the outgoing null geodesic equation which is given by,
\begin{equation}
\frac{dt}{dr}=e^{\psi-\nu}
\label{eq:null1}
\end{equation}
We use a method which is similar to that given in 
\cite{kn:dim}. 
The singularity curve is given by $v(t_s(r),r)=0$, which 
corresponds to $R(t_s(r),r)=0$. Therefore, if we have any future directed 
outgoing null geodesics terminating in the past at the singularity, 
we must have $R\rightarrow 0$ as $t\rightarrow t_s$ along the same. 
Now writing \e(\ref{eq:null1}) explicitly in terms of variables 
$(u=r^\alpha,R)$, we have,
\begin{equation}
\frac{dR}{du}=\frac{1}{\alpha}r^{-(\alpha-1)}R'\left[1-\sqrt
{\frac{be^{2\nu_{0}}+\frac{r^{3}\mathcal{M}}{R}+\frac{\Lambda R^{2}}
{3}-1}{be^{2\nu_{0}}}}\,\right]
\label{eq:null2}
\end{equation}
\e (\ref{eq:null2}) is required to be {\it finite and positive}, 
for the existence of a naked singularity
\cite{kn:dim}. 
In order to get the tangent to the null geodesic in the $(R,u)$ plane, 
we choose a particular value of $\alpha$ such that the geodesic equation 
is expressed only in terms of $\left(\frac{R}{u}\right)$. A specific 
value of $\alpha$ is to be chosen which enables us to calculate 
the proper limits at the central singularity. In the tangential pressure 
collapse model discussed in the previous section we have $\X_1(0)=0$, 
and hence we choose $\alpha=\frac{7}{3}$ so that when the limit 
$r\rightarrow 0,t\rightarrow t_{s}$ is taken we get the value of tangent 
to null geodesic in the $(R,u)$ plane as,
\be
\frac{dR}{du}=\frac{3}{7}\left(\frac{R}{u}+\frac{\sqrt{M_{0}}\X_{2}(0)}
{\sqrt{\frac{R}{u}}}\right)\frac{(1-\frac{F}{R})}{\sqrt{G}(\sqrt{G} 
+\sqrt{H})}
\label{eq:null4}
\eq

Now note that for any point with $r>0$ on the singularity 
curve $t_s(r)$ we have $R\to 0$ whereas, $F$ (interpreted as the 
gravitational mass within the comoving radius $r$) tends to a finite 
positive value once the energy conditions are satisfied. Under 
the situation, the term $F/R$ diverges in the above equation, and all 
such points on the singularity curve will be covered as there will 
be no outgoing null geodesics from such points.

We hence need to examine the central singularity at $r=0,R=0$ to 
determine if it is visible or not. That is, we need to determine if 
there are any solutions existing to the outgoing null geodesic equation, 
which terminate in the past at the singularity and in future go to a 
distant observer in the spacetime, and if so under what conditions these 
exist. Let $x_{0}$ 
be the tangent to the null geodesics in $(R,u)$ plane, at the central 
singularity, then it is given by,
\begin{equation}
x_0=\lim_{t\rightarrow t_s}\lim_{r\rightarrow 0} 
\frac{R}{u}=\left.\frac{dR}{du}\right|_{t\rightarrow t_s;r\rightarrow 0}
\end{equation}
Using \e (\ref{eq:null4}), we get,
\begin{equation}
x_0^{\frac{3}{2}}=\frac{7}{4}\sqrt{\M_0}\X_2(0)
\end{equation}
In the $(R,u)$ plane, the null geodesic equation will be,
\be
R=x_0u
\eq
while in the $(t,r)$ plane, the null geodesic equation near the 
singularity will be,
\begin{equation}
t-t_s(0)=x_0r^{\frac{7}{3}}
\end{equation}

It follows that if $\X_2(0)>0$, then that implies that 
$x_0>0$, and we then have radially outgoing null geodesics coming 
out from the singularity, making the {\it central singularity locally 
visible}. On the other hand, if $\X_2(0)<0$, we will have a {\it black 
hole solution}. We have, however, already seen in \e (\ref {eq:ttan}), 
that the value of $\X_2(0)$ entirely depends upon the initial data 
and the cosmological term $\Lambda$. Given any $\Lambda$, the initial data 
can always be chosen 
such that the end state of the collapse would be either a naked 
singularity or a black hole. Hence, it follows that  
{\it for both positive and negative cosmological constants, a naked 
singularity can occur as the final end state of gravitational collapse}.

We noted here earlier that dust collapse models have been 
analyzed in the presence of a cosmological constant.
The nature of the final singularity in that case has been 
analyzed using the so called `roots' method 
(Deshingkar S. S. {\it et al}
\cite{kn:psj1}). We show below that the different treatment we 
have used above to deal with the collapsing clouds with pressure
included, arrives at similar conclusions when specialized to the
dust case.

The Einstein equations for {\it dust} may be obtained by putting 
$p_{\theta}=0$ in the tangential pressure fluid model above. 
They take the form,
\begin{equation}
\rho(r,t)= \frac{\mathcal{M}_{d}^{'}(r)}{R^{2}R^{'}}
\end{equation}
\begin{equation}
\dot R^{2}=\frac{\mathcal{M}_{d}(r)}{R}+f(r)+\frac{\Lambda R^{2}}{3}
\end{equation}
where $f(r)=r^{2}b_{0}(r)$ is the dust {\it energy free function}. 
Since we are interested in collapse we must have~$\dot R\leq$0. 
Then we have from the above equations,
\begin{equation}
\sqrt{R}\dot R= - \sqrt{\mathcal{M}_{d}(r)+f(r)R+\frac{\Lambda R^{3}}{3}}
\label{eq:ddyn}
\end{equation}
Using the scaling freedom we may write,
\begin{equation}
\mathcal{M}_{d}(r)=\int_{0}^{r} \rho(0,r) r^{2} dr=\rho^{\tiny{avg}}(r)  r^{3}
\end{equation}
Rewriting equation (\ref{eq:ddyn}) in terms of the scale factor $v$,
\begin{equation}
\sqrt{v}\dot v= - \sqrt{\rho^{\tiny{avg}}(r) +\frac {f(r)v}{r^{2}}
+\frac{\Lambda v^{3}}{3}}
\label{eq:ddyn1}
\end{equation}
Define $\frac {f(r)v}{r^{2}}$ as $L(r,v)$ which is at least  
a $C^{2}$ function of its arguments. Then $t_{s}(r,v)$ is readily 
calculated from (\ref{eq:ddyn1}) to be,
\begin{equation}
t_{s}(r,v) = \int_{v}^{1}\frac{\sqrt{v} dv }
{\sqrt{\rho^{\tiny{avg}}(r)+L(v,r)+\frac{\Lambda v^{3}}{3}}}
\end{equation}
The shell focusing singularity $R=0$ occurs first for the comoving 
coordinate $r=0$. The time of its formation is,
\begin{equation}
t_{s}(0) = \int_{0}^{1}\frac{\sqrt{v} dv }{\sqrt{\rho^{\tiny{avg}}(0)
+L(v,0)+\frac{\Lambda v^{3}}{3}}}
\end{equation}
As we did for the fluid model, we Taylor expand $t_{s}(r)$ near $r=0$
to get,
\begin{equation}
t_{s}(r)=t_{s}(0)+r\mathcal{X}_{1}(0)+r^{2}\mathcal{X}_{2}(0)+\mathcal{O}(r^{3})
\end{equation}
where $\X_{1}(v)=0$ (assuming smooth initial data), and
\begin{equation}
\X_{2}(v)=\,\,-\int_{v}^{1}\frac{B_{d}^{''}(0,v)\sqrt{v}\,dv }
{B_{d}(0,v)^{3/2}}
\label{eq:ttan1}
\end{equation}
with $B_{d}(r,v)={\rho^{\tiny{avg}}(r)+L(v,r)+\frac{\Lambda v^{3}}{3}}$. 
Again, it needs to be analyzed whether there exists future directed 
null geodesics that end at the singularity in the past. Consider the 
marginally bound case $(f(r)=0)$. Then the equation of the null 
geodesic is,
\begin{equation}
\frac{dt}{dr}= R^{'}
\label{eq:dng}
\end{equation}   
In terms of $(u,r)$ it is written as,
\begin{equation}
\frac{dR}{du}=\frac{R^{'}}{\alpha r^{\alpha-1}} 
(1-\sqrt{\frac{\mathcal{M}_{d}(r)}{R}+\frac{\Lambda R^{2}}{3}})
\label{eq:dng1}
\end{equation}
Choosing $\alpha=\frac{7}{3}$,
\begin{equation}
\frac{dR}{du}=\frac{3}{7}(\frac{\sqrt{v}v^{'}}{\sqrt{\frac{R}{
u}}}+\frac{R}{
u})
 (1-\sqrt{\frac{\mathcal{M}_{d}(r)}{R}+\frac{\Lambda R^{2}}{3}})
\end{equation}

Taking the limits as before, the final expression becomes,
\begin{equation}
\frac{4}{7}\, x_{0}^{\frac{3}{2}}=\sqrt{\rho^{avg}_{0}}\mathcal{X}_{2}(0)
\label{eq:dtan}
\end{equation}
It follows that for both positive and negative 
values of $\Lambda$, there exists initial data that may give  
$\mathcal{X}_{2}(0)>0$. Thus the dust central singularity may 
be locally visible, even when there is a non-zero cosmological constant
present. It may thus be claimed that the dust central naked singularity 
is not precluded by $\Lambda$. 
\end{section}

\begin{section}{Conclusions}

The final outcome of gravitational collapse is one of the most
important open problems in gravitation theory, and the study of the
fluid with vanishing radial pressure, and also the dust collapse models here indicate that the presence of a non-vanishing $\Lambda$ cannot, in any conclusive
manner, act as a cosmic censor. Although the value of the null geodesic
tangents are modified by the presence of $\Lambda$, there are still cases
where the singularity is locally visible depending on the initial data.
The global visibility of such a singularity (null rays from the
singularity reaching an asymptotic observer) which is locally naked, will
depend on the overall behaviour of the various functions concerned
which appear in the analysis. But as such general relativity has no natural
length scale attached to it, and so a locally visible singularity may  
violate the cosmic censorship as effectively as a globally visible
singularity. Moreover, it is seen from various examples that if a
singularity is locally naked, then one can always make it globally
visible by suitable choice of the allowed functions, or by restricting
properly the boundary of the cloud
(\cite{kn:gasp}).
We also note that studies pertaining to radiation collapse (of a   
{\it Type II} matter field
\cite{kn:hwe})
in spacetimes with $\Lambda$, exist in literature
\cite{kn:lemos}.
Such studies also support the conclusion that both collapse
endstates, namely a black hole or a naked singularity are
possible in the presence of a non-zero $\Lambda$.

The currently observed accelerated expansion of the universe, 
and also the AdS/CFT conjecture in string theories have generated 
considerable interest in spacetime scenarios with a cosmological constant. 
We studied here dynamical collapse of a tangential pressure fluid model 
with non-vanishing $\Lambda$. Various aspects of gravitational collapse 
with $\Lambda$ have been discussed in the context of the fluid model 
which include dynamics, junction conditions, apparent horizons, and 
nature of the singularity. The question of cosmic censorship in the 
collapse model is discussed, and it is also shown that when the pressures 
are put to zero in the model it reduces to the dust case.

The presence of a finite cosmological constant, it is seen, affects 
the dynamical collapse of the fluid in many ways. To summarize,

1. $\Lambda$ plays a role in deciding the dynamical evolution of a shell. 
The conditions for singularity formation and bounce are derived.

2. The matching conditions for the interior and exterior spacetimes 
were discussed, assuming that the exterior is asymptotically AdS or dS.

3. The apparent horizon conditions and the time of horizon formation 
were studied in brief for $\Lambda\neq0$.

4. The effect of $\Lambda$ on the nature of the central singularity 
was analyzed. The condition for local visibility was derived and it was 
seen that even for non-vanishing $\Lambda$ both black hole and naked 
singularity are possible end states of collapse. 
\end{section}

{\bf Acknowledgments\\}
A.M. would like to acknowledge support from TIFR, Mumbai during the 
Visiting Students Research Programme 2004 and thanks Ashutosh Mahajan for 
discussions.

\end{document}